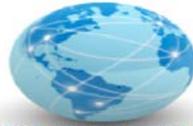

# The Chromospheric Magnetic Field in Active Regions Derived from Spectropolarimetry of Ca II 8542

**Ali G. A. Abdelkawy**
Department of Astronomy and Meteorology, Faculty of Science, Al-Azhar University, Nasr City, Cairo, Egypt, Postal Code 11884.
ali.astro@azhar.edu.eg

**Abdelrazek M. K. Shaltout**
Department of Astronomy and Meteorology, Faculty of Science, Al-Azhar University, Nasr City, Cairo, Egypt, Postal Code 11884.
ahaltout@azhar.edu.eg

**M. M. Beheary**
Department of Astronomy and Meteorology, Faculty of Science, Al-Azhar University, Nasr City, Cairo, Egypt, Postal Code 11884.
mmbeheary2007@yahoo.com

**ABSTRACT**

We analyze spectropolarimetric observations of the chromospheric Ca II 8542 line taken by the Interferometric Bidimensional Spectrometer (IBIS) at the Dunn Solar Telescope and photospheric Fe I 6302 lines obtained with the Solar Optical Telescope (SOT) of Spectro-polarimeter (SP) on board the Hinode satellite. The data were obtained on 2012 January 29 targeting NOAA active region (AR) 11408 and AR 11410. By using the center-of-gravity (COG) method we compute the line-of-sight (LOS) field strength for observed lines of Fe I 6301.5/6302.5 with SP spacecraft at every places in the active region. Also, we construct a COG map for the chromospheric IBIS data of Ca II 8542 line in comparison with COG maps derived from the SP of field strength. We found the photospheric field strength ranges up to 2 kG in plages and up to 2 kG or higher values inside the umbral region. In the case of chromospheric field strength, the LOS field inside the umbral region increases up to 800 G, and the field strength decreases toward the edges of sunspot.

## 1. INTRODUCTION

The IBIS is a two-dimensional spectro-polarimeter that has been installed for the polarimetry of the Sun (Cavallini 2006; Reardon & Cavallini 2008) established at the Dunn Solar Telescope. It produces polarimetric data of the four Stokes IQUV profiles using for measurements of the magnetic field on the Sun. The spectropolarimetric data allows us a best possibility to derive all components of the field strength with high spatial resolution, as recently determined by (Ichimoto & Shaltout 2012; Shaltout & Ichimoto 2015).

The Solar Optical Telescope (SOT) on board the Hinode Kosugi et al. (2007) satellite has provided two kinds of spectropolarimetric data of spectral line profiles of full Stokes taken by Spectro-polarimeter and Stokes I and V images obtained by the Narrowband Filter Imager (NFI). Since the SOT/SP data have a spatial resolution of 0.3 arcsec per pixel. The high spatial resolution of SP observations of sunspots show that there are a lot of different morphological phenomena of dark central regions, called umbra and a surrounding less dark filamentary region called penumbra.

The COG is a simple approach to infer the LOS field strength from the observed Stokes I and V profiles without taking into account the filling factor of magnetic elements. It has been adopted in various investigations (Rees & Semel 1979; Cauzzi et al. 1993; Uitenbroek 2003; Balasubramaniam et al. 2004; Nagaraju et al. 2008; Kleint et al. 2009). The COG gives information on the polarity of field, weak and strong field regions, and it can be used for the comparison between photospheric and chromospheric magnetic fields.

In the current paper we employ the COG method described by Rees & Semel (1979) to obtain the magnetic field strength from the observational of chromospheric and photospheric data within an accuracy of about 10%. Given the advantage of the high-resolution observations, the COG can be applied without using model of radiative transfer. The observations are explained in section 2. The center-of-gravity and results are described in section 3. Finally, a conclusion is presented in section 4.

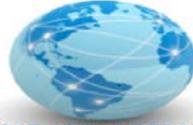

## 2. OBSERVATIONS

The sunspot of NOAA 11408 was observed on 2012 January 29 with the IBIS, which has been provided by Schad (2015). It was located at a heliocentric angle of ($\mu = \cos\theta = 0.8$) at the time of observation. The IBIS recorded full Stokes profiles of Ca II line at 8542.1 Å (effective Landé factor g = 1.10) with a spatial sampling of 0″.0976. The observation began at 16:33 UT and finished at 17:04 UT. The spectral sampling of Ca II 8542 was covered a range from 8540.35 to 8542.85 Å with an interval of 2.5 Å over the whole line profile. In full Stokes mode, six modulated polarization states were measured (I+Q, I+V, I-Q, I-V, I-U, I+U) at each spectral sample. The wavelength points across the spectral line was about 20 with a cadence of 50 sec. The IBIS image has a field of view (FOV) of 45″ x 95″.

The SP data was observed on 2012 January 29 for determining the LOS field strength in active region. The target active region analyzed here is AR 11410 located at (X= -562″, Y= 378″) at the observed time. The SP listed full Stokes profiles of Fe I lines at 6301.5 Å (effective Landé factor g = 1.67) and 6302.5 Å (g = 2.5). The Stokes IQUV spectral profiles were obtained with an integration time of 3.9 s at each step.

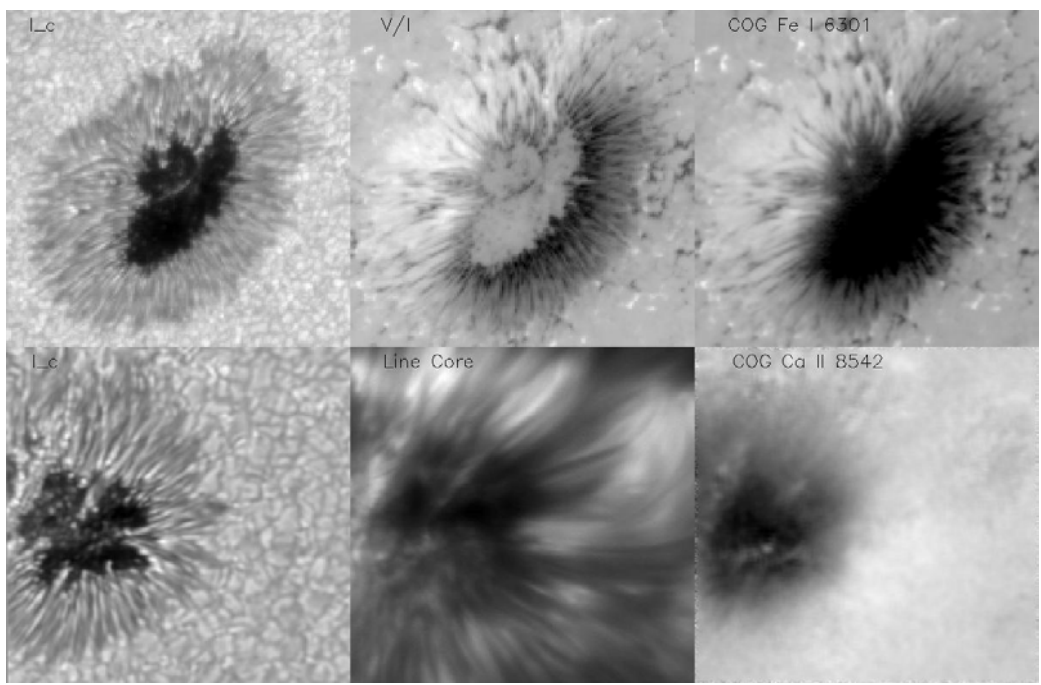

**Fig 1:** Images exhibit two sunspots of AR 11408 and AR 11410 observed on 2012 January 29, which are analyzed in the current work. Top row, left to right: continuum intensity ($I_c$), V/I and COG maps of Fe I 6301.5 line of SP data, respectively. Bottom row, based on order of images from left to right: continuum intensity ($I_c$), line core and COG maps of the IBIS data. The COG image of IBIS is scaled over the range of -1.5 kG to 0.5 kG, while the COG map of SP is shown over ranging from -3 kG to 3 kG.

The observations started at 22:01 UT and ended at 23:04 UT as a fast raster scan mode. The SP image took about 60 minutes with 1000 slit steps to map the FOV. The raster scan covered a FOV of 295″ x 162″. We used the IDL standard routine, SP_PREP (Lites & Ichimoto 2013), available under Solar Software (SSW), to calibrate the SP data. Since the step size was set to 0″.295 and the number of pixels along the slit was 512. The pixel size along the slit direction was 0″.317.

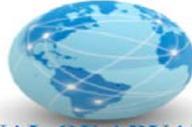

## 3. CENTER-OF-GRAVITY (COG) METHOD AND RESULTS

From the IBIS and SP data, we use the observed Stokes I and V profiles for given wavelengths with the COG method, as has been described in Nagaraju et al. (2008). Since the LOS magnetic field (in Gauss) was obtained by the formula:

$$B_{LOS} = \frac{(\lambda_+ - \lambda_-)/2}{4.67 \times 10^{-13} \lambda_0^2 g_L} \quad (1)$$

where $g_L$ is the effective Landé factor and ($\lambda_0$) is the central wavelength of the line in angstrom. The COG wavelengths ($\lambda_\pm$) of the positive and negative circularly polarized components are defined as:

$$\lambda_\pm = \frac{\sum \lambda[1 - (I \pm V)/I_c]}{\sum [1 - (I \pm V)/I_c]} \quad (2)$$

where $I_c$ is the local continuum intensity. Finally, we used this formula in our analysis:

$$B_{LOS} = \alpha(\lambda_+ - \lambda_-) \quad (3)$$

Since the coefficient α is computed from:

$$\alpha = 1/[2 \times 4.67 \times 10^{-13} g_L \lambda_0^2] \quad (4)$$

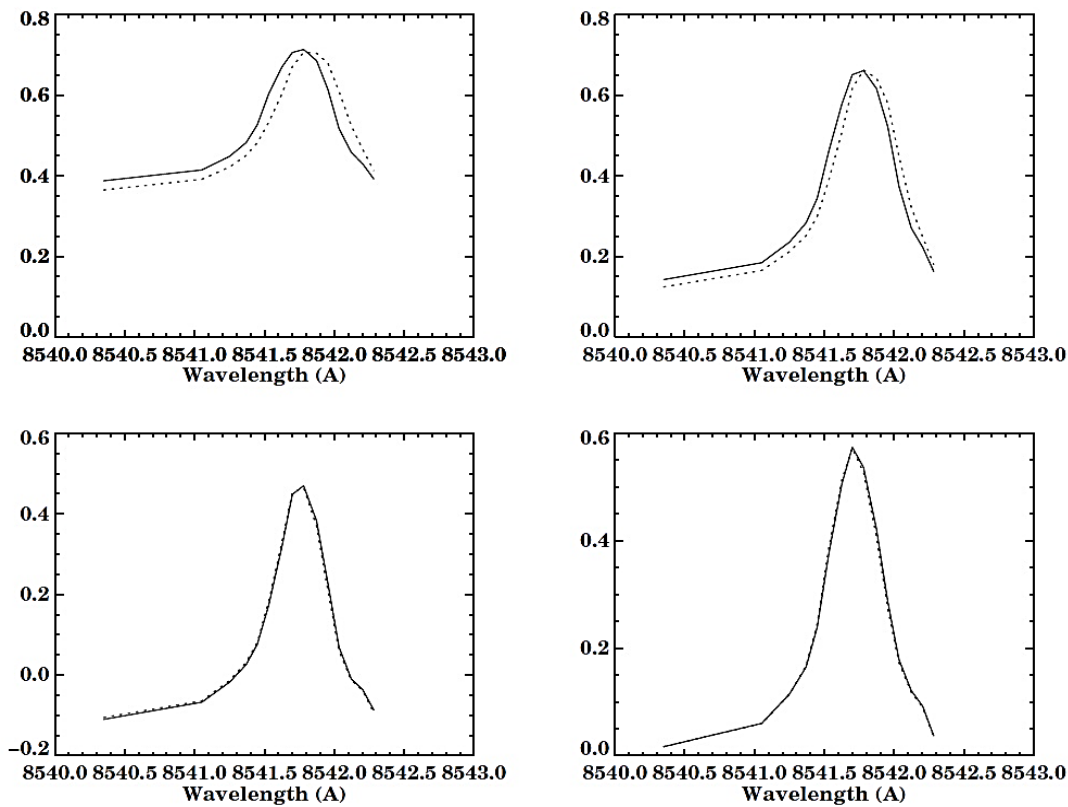

Fig 2: Plots of positive circularly polarized component $[1 - (I + V)/I_c]$ (solid line) and negative circularly polarized component $[1 - (I - V)/I_c]$ (dotted line) of Ca II 8542 line at different positions in active region: umbra, penumbra and two positions outside sunspot regions. The LOS magnetic fields derived from the four locations are -924 G, -526 G, 102 G and 58 G, respectively.

Figure 1 displays the continuum intensity ($I_c$), V/I and $B_{LOS}$ maps of SP observations found at the top row from left to right, while the IBIS data, includes the white light (I), line core of Ca II 8542 and $B_{LOS}$ images, which presented at the bottom row of this figure.

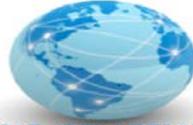

The coefficient α value of Ca II 8542 line is 13.3 kG/Å but its value is 10.8 kG/Å for the Fe I 6302.5 line and 16.1 kG/Å for the Fe I 6301.5 line. The integration (summation) is over the spectral range of a given spectral line. Figure 2 shows the COG wavelengths selected at four places in active region for the illustration, as computed from Equation 2 with the observed IBIS data.

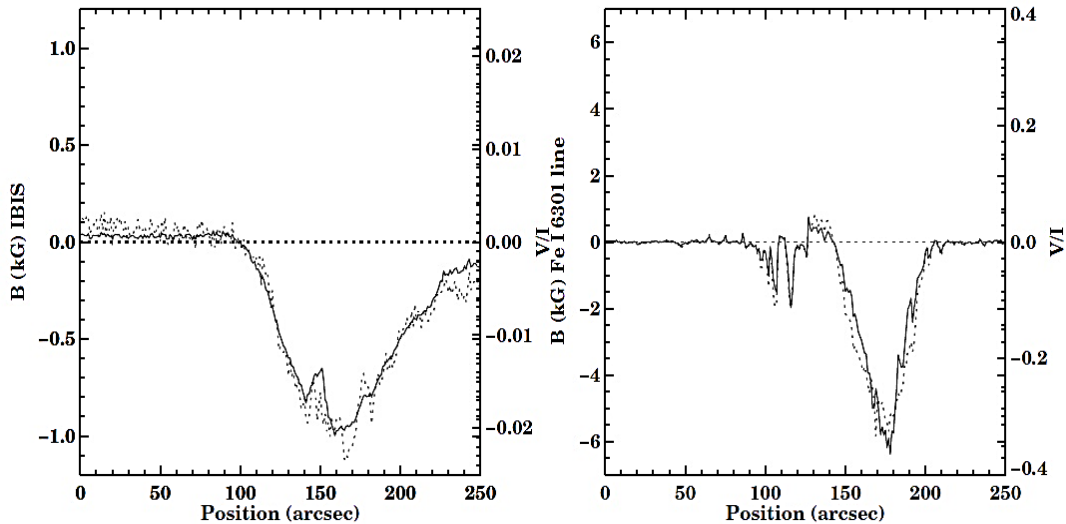

**Fig 3:** Plots of V/I profiles (dotted line) and LOS field strengths (thick solid line) at the photosphere (right) and chromosphere (left) along two radial cuts plotted on SP and IBIS images selected from south-north directions passing through the center of the sunspot.

We notice that the blending in the spectral I profiles is less serious, therefore, we did not use cleaning processes as a linear combination of Gaussian functions, because the shapes of Stokes I and V profiles look familiar in the umbral regions of sunspot, as shown in figure 2 at the top row of left panel. The plots of photospheric and chromospheric LOS field strengths along two radial cuts are exhibited in figure 3. Using the SP data, the right side of figure 3 shows the plot of V/I profile versus the photospheric LOS magnetic field along a radial slice through the central umbra of the sunspot image. The photospheric field strength along this radial cut increases up to about 2000 G, which is consistent with the result suggested by Chae et al. (2007). With the IBIS data, the V/I profile against the chromospheric LOS field strength is presented on the left side of figure 3. We take the same position on the sunspot image of IBIS, as done with SP data. The chromospheric field strength increases up to about 800 G, our result is in agreement with that of Nagaraju et al. (2008).

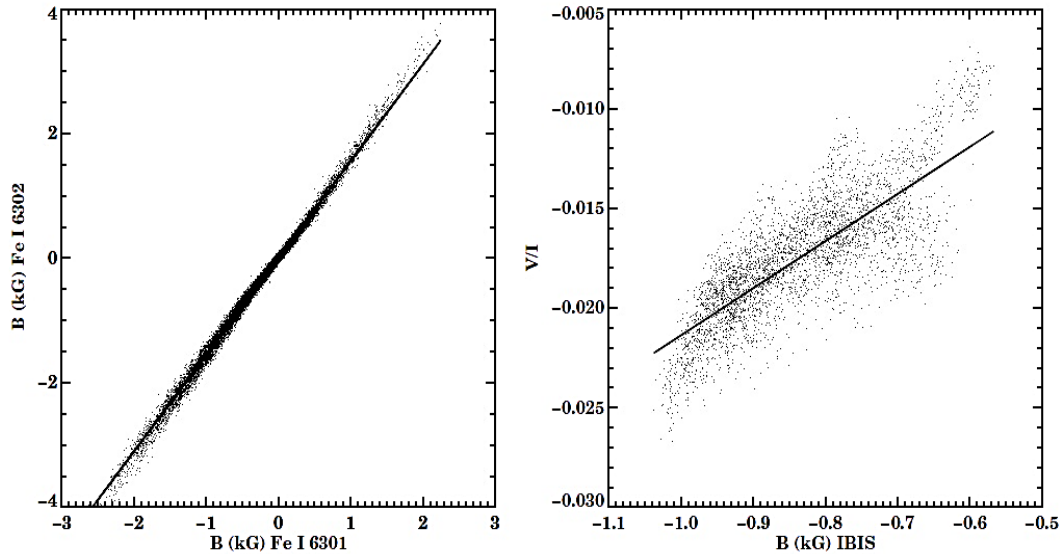

**Fig 4:** Right: Plot of V/I profile along the chromospheric LOS magnetic field determined from the center-of-gravity result in the umbral region of sunspot. Left: LOS magnetic field strengths of Fe I 6302 lines in regions outside sunspot. The solid line refers to the line best fit.

The photospheric field strength is larger in the umbral region and decreases toward the edges of the sunspot. In the case of chromospheric field strength, the same features were noticed, but the chromospheric field values are considerably smaller as compared to the photospheric field.

In the right panel of figure 4, we show the V/I profile along the chromospheric field strength in the umbral region of sunspot. The solid line represents the line fit through these points. This figure indicates a coherent result of magnetic fields in the umbra, showing that a linear relationship between V/I and $B_{LOS}$ results. Also, the correlation coefficient between V/I and $B_{LOS}$ is found to be 0.79 at umbral regions. Figure 4 (left panel) displays LOS magnetic field strengths of Fe I 6301.5/6302.5 lines in regions outside sunspot, which obtained with the SP observation. The correlation between V/I and $B_{LOS}$ of Fe I 6301 is 0.98 at the regions outside sunspots. The main features at regions outside sunspots are plages and networks. Our result exhibits that these features have LOS fields ranged from –2 kG to 2 kG, and the same result was verified by Chae et al. (2007).

## 4. CONCLUSION

In this paper, we analyzed the chromospheric LOS component of magnetic field of Ca II 8542 taken by the IBIS instrument and photospheric Fe I 6301.5/6302.5 observed on board the Hinode satellite, respectively.

We adopted the COG approach proposed by Rees & Semel (1979) to derive the magnetic field strengths, showing that the chromospheric LOS field strength is from -1.5 kG to 0.5 kG, while the photospheric LOS field strength increases up to 2 kG.

We compare the photospheric and chromospheric LOS magnetic field strengths in two active regions. The chromospheric magnetic field strength is weaker as compared to the photospheric field, as has been confirmed by (Solanki 2003; Nagaraju et al. 2008). It is interesting that this analysis was obtained without taking into account the filling factor of magnetic elements.

The photospheric LOS field strengths outside sunspot regions are scaled over ranges of –2 kG to 2 kG, therefore, our finding was supported by Chae et al. (2007).

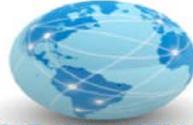

**GLOBAL JOURNAL OF ADVANCED RESEARCH**
(Scholarly Peer Review Publishing System)

## 5. ACKNOWLEDGMENTS

This work was supported by the Faculty of Science, Al-Azhar University funded by the Egyptian Government. Hinode is a Japanese mission developed and launched by ISAS/JAXA, with NAOJ as a domestic partner and NASA and STFC (UK) as international partners. It is operated by these agencies in co-operation with ESA the NSC (Norway). The National Solar Observatory in (U.S.A) is operated by the Association of Universities for Research in Astronomy, Inc., under cooperative agreement with the National Science Foundation. IBIS is a project of INAF/OAA with additional contributions from University of Florence and Rome and NSO. Gianna Cauzzi of National Institute of Astrophysics, Rome, has helped for getting the IBIS data, which is much appreciated by the authors. We would thank K. Ichimoto for the participation of Al-Azhar University in the analysis of Hinode data based on a Letter of Agreement signed in 2011.

## 6. REFERENCES

[1] Balasubramaniam, K. S. et al. 2004, ApJ, 606, 1233

[2] Cauzzi, G. et al. 1993, Sol. Phys., 146, 207

[3] Cavallini, F. 2006, Sol. Phys., 236, 415

[4] Chae, Jongchul, et al. 2007, PASJ, 59, 619

[5] Ichimoto, K. & Shaltout, A. M. 2012, 39th COSPAR Scientific Assembly. Held 14-22 July 2012, in Mysore, India, Abstract B0.1-39-12,P.789

[6] Kleint, L., Reardon, K., Stenflo, J. O., Uitenbroek, H. and Tritschler, A. 2009, June. In Solar Polarization 5: In Honor of Jan Stenflo (Vol, 405, 247)

[7] Kosugi, T. et al. 2007, Sol. Phys., 243, 3

[8] Lites, B. W. & Ichimoto, K. 2013, Sol. Phys., 283, 601

[9] Nagaraju, K. et al. 2008, ApJ, 678, 531

[10] Reardon, K. P. & Cavallini, F. 2008, A&A, 481, 897

[11] Rees, D. E. & Semel, M. D. 1979, A&A, 74, 1

[12] Schad, T. A. 2015, private communication

[13] Shaltout, A. M. K. & Ichimoto, K. 2015, PASJ, 67, 27

[14] Solanki, S. K. 2003, The Astron. Astrophys. Rev., 11, 153

[15] Uitenbroek, H. 2003, ApJ, 592, 1225